\documentclass[fleqn,twoside]{article}
\usepackage{espcrc2}
\usepackage{graphicx}
\usepackage{epsf}

\usepackage{amsmath, amssymb,latexsym,float,algorithm,algorithmic}
\usepackage{enumerate}

\newcommand{\R}{\text{\fontshape{n}\selectfont I\kern-.42exR}}

\newcommand{\1}{\text{\fontshape{n}\selectfont 1\kern-.56exl}}

\title{Global Monte Carlo for fermions using ordered statistics}

\author{A. Bori\c{c}i
       \address{University of Edinburgh, Department of Physics and 
Astronomy \\
                Mayfield Rd, JCMB, Edinburgh EH9 3JZ}
       }

\begin{document}

\begin{abstract}
In this talk I discuss a new possibility for stochastic representation of the fermion determinant. The method can be used for global Monte Carlo fermion algorithms and is tested in the case of the Schwinger model.
\vspace{1pc}
\end{abstract}

\maketitle

\section{INTRODUCTION}

The standard method for stochastic representation of the fermion determinant is
the pseudofermion method \cite{WP}. It allows to trade the computation of the
determinant in favour of extra noise in the path integral of lattice QCD.
The other alternative is to use the stochastic Taylor expansion of \cite{BK}.
This suffers from the so called probability sign problem which is created
by the noise \cite{KK}.
The problem is less apparent if the method is applied to a
fractional power of the determinant, a fact which is used by the Kentucky
Monte Carlo algorithm \cite{KMC}.

In this talk I show that the sign problem created by the noise can be
eliminated altogether if one uses a certain statistic from a sample of unbiased
noisy estimators of the effective fermion action. As expected, a statistics which is an
unbiased estimation of the effective fermion action cannot yield an
unbiased estimation of the fermion determinant. The opposite logic suggests that a
biased estimation of the former could lead in principle to an unbiased
estimation of the latter if this is done ''correctly''. This is indeed the
case if one chooses a certain order statistic from the sample. I show below how
this can be realized using the central limit theorem and the asymptotic distribution of a central order statistic.

\section{USING THE CENTRAL LIMIT THEOREM}

Let $X_1,X_2,\ldots,X_n$ be unbiased noisy estimators of the fermion
effective action $S_{\text{eff}} = - \text{Tr log} (D^{\dag}D)$, where $D$ is a
generic Dirac. A particular way (and popular) to do it is using real (there is
nothing special, one can use complex numbers as well) random
vectors $Z_1,Z_2,\ldots,Z_n$ such that:
\begin{equation}
X_i = - Z_i^{\dag} [\text{log} (D^{\dag}D)] Z_i 
\end{equation}
Computation of the right hand side can be done for example using Lanczos methods
as described in detail in \cite{Bai_et_al,Borici_UVSFb}. It is easy to show that
$X_i$ is an unbiased estimator of $S_{\text{eff}}$. Another useful result (that
so far went unappreciated and wasn't used)
is the normality of the $X_i$ distribution. This can be
easily shown for $Z_2$ noise vectors using the central limit theorem. Note that
the whole idea relies on this result.   

\section{BASICS OF ORDER STATISTICS}

Everyone has encountered at least once the notion of median as an
unbiased estimator of the population mean. It is identified as the middle point
after the sample of real numbers $X_1, X_2, \ldots , X_n$ drawn from the same
continuous (to fix the idea)
distribution $f(x), x \in R$ has been ordered in the ascending order:
\begin{equation}
X_{(1)}, X_{(2)}, \ldots , X_{(n)}
\end{equation}
This is possible if the sample size $n$ is odd. In case $n$ is even the median
is defined to be any number in the interval $[X_{(n/2)}, X_{(n/2+1)}]$.

The median is a statistic as it is the arithmetic average. It is nothing
particular to the median as an instance of a statistic calculated from the
ordered data. In general, $X_{(k)}, k = 1, \ldots, n$ is called the $k$'th
{\it ordered statistic}.

{\bf Distribution} of $X_{(k)}$ can be found using a simple heuristic. First,
let $f_k(x)dx$ be the probability that $X_{(k)}$ is in an infinitesimal interval
$dx$ about $x$. This means that one of the sample variables is in this interval,
$k-1$ are less than $x$ and $n-k$ are greater than $x$. The number of ways of
choosing these variables is the multinomial coefficient $C(n;k-1,1,n-k)$. If
$F(x)$ is the cumulative distribution function of the sample variables then one
gets:
\begin{equation}
f_k(x) dx = [F(x)]^{k-1} [1 - F(x)]^{n-k} f(x) dx
\end{equation}
This result can be formally proven. As a starting classic is the book of
H.A. David \cite{David}. A recent introduction to the subject is provided in
\cite{ABN}.

\section{ASYMPTOTIC THEORY}

The distribution of an order statistic can be of little use for direct
analytical calculations of moments for most of known parent distributions.
Of course one can use numerical methods to compute numerical values to the
desired accuracy. However, as the sample size grows the order statistic
distribution approaches some limiting distributions. The situation is analogous
to the central limit theorem. In general the distribution is not necessarily
asymptotically normal as one could have expected although this will be the case
for the {\it central order statistics} that will be dealt with below.

For $0 < \alpha < 1$, let $i = [n\alpha] + 1$, where $[n\alpha]$
represents the integer part
of $n\alpha$. Let also $F$ be absolutely continuous with $f$ positive at
$F^{-1}(\alpha)$ and continuous at that point.
Then the following result applies:

{\bf Theorem} (asymptotic distribution of a central order statistic).
As $n \rightarrow \infty$,
\begin{equation}
\sqrt{n} f[F^{-1}(\alpha)] \frac{X_{(i)} - F^{-1}(\alpha) }
{ \sqrt{\alpha(1-\alpha)} } \rightarrow N(0,1)
\end{equation}
The proof can be found eg, in \cite{ABN}.

{\it Example}. Suppose that the distribution is symmetric around the population
mean $\mu$. Let also assume that the variance $\sigma^2$ is finite and $f(\mu)$
is finite and positive. For simplicity $n$ is taken to be odd. Then the median
$X_{(n+1)/2}$ is an unbiased estimator of $\mu$ and asymptotically normal.
Further, Var$[X_{(n+1)/2}] \approx [4nf(\mu)^2]^{-1}$.

\section{UNBIASED ESTIMATION OF THE FERMION DETERMINANT}

Let the $X_i, i=1,...,n$ be a normally distributed sample of unbiased estimators
of the fermion effective action.
I am interested in the expected value of $e^{-X_{(i)}}$ with $i = [n\alpha] + 1$
and $\alpha$ as above. Let $X \sim N(\mu,\sigma)$. It is easy to show that:
\begin{equation}
<e^{-X}> = e^{-\mu + \sigma^2/2}
\end{equation}
Hence, an unbiased estimation of the effective action gives a biased estimation
of the fermion determinant.
Now let $x_{\alpha} := (F^{-1}(\alpha) - \mu)/\sigma$.
Then $X_{(i)}$ is asymptotically normal with mean $\mu + x_{\alpha}\sigma$
and variance $2\pi\sigma^2\alpha(1-\alpha)e^{x_{\alpha}^2}/n$. For the
ordered statistic one gets:
\begin{equation}
<e^{-X_{(i)}}> = e^{-\mu - x_{\alpha}\sigma +
\pi\sigma^2\alpha(1-\alpha)e^{x_{\alpha}^2}/n}
\end{equation}

To have an unbiased estimator of the fermion determinant in terms of an
ordered statistic one should have $<e^{-X_{(i)}}> = e^{-\mu}$. This can be
achieved if:
\begin{equation}
x_{\alpha} = \pi\sigma\alpha(1-\alpha)e^{x_{\alpha}^2}/n
\end{equation}

The crucial question here is: given $n$ and $\sigma$ is there any solution of
the above equality in terms of $x_{\alpha}$? To answer this question I express first $\alpha$ in terms of $x_{\alpha}$. Since $(X - \mu)/\sigma \sim N(0,1)$ then I have:
\begin{equation}
\alpha = P(X < x_\alpha) = \frac{1}{2} + \frac{1}{2}
\text{erf}(\frac{x_{\alpha}}{\sqrt{2}})
\end{equation}
Hence, the equation to be solved is:
\begin{equation}
x_{\alpha} = \frac{\pi\sigma}{4n}
[1-\text{erf}(\frac{x_{\alpha}}{\sqrt{2}})^2]e^{x_{\alpha}^2}
\end{equation}
Using Taylor expansion around $x_{\alpha} = 0$ one gets:
\begin{equation}
x_{\alpha} = \frac{\pi\sigma}{4n}[1 + O(x_{\alpha}^2)]
\end{equation}
For small $x_{\alpha}$ the $O(x_{\alpha}^2)$ can be neglected one gets a unique
solution:
\begin{equation}
x_{\alpha} = \frac{\pi\sigma}{4n}, ~~~~~~x_{\alpha} << 1
\end{equation}
In this case the sample size should be much greater then $\sigma$ and $i$ can be
calculated to give:
\begin{equation}
i = [n\alpha] + 1 = [\frac{n}{2} + \sqrt{\frac{\pi}{2}} \frac{\sigma}{4}] + 1 
\end{equation}

Thus, using order statistics it is possible to have an unbiased estimator of the
fermion determinant from a biased high estimator of the fermion effective
action.

However, the estimator is only asymptotically unbiased meaning that the sample
size has to be large enough. In practice, at least from the examples below the
asymptotic regime is reached for $n = 40$. In many application to lattice
fermions $\sigma$ is large anyway. Thus, requiring $n >> \sigma$ will satisfy
the conditions of the asymptotic normality as well.

Note also that $\sigma$ is unknown and has to be estimated from the sample.
This is a source of systematic error that may bias the stochastic estimation
of the determinant. To compute the error one has to substitute
$\sigma$ with its estimator $S$ in the formula for $x_{\alpha}$.
Then the average error is given by:
\begin{equation}
<e^{- x_{\alpha}\sigma + \frac{\pi\sigma^2}{4n}}> =
<e^{\frac{\pi\sigma}{4n}(\sigma - S)}>
\end{equation}
The error has to be computed in order to control the systematic errors.
This can be done using for $\sigma$ large samples to guarantee small errors
and then substitute in the above formula.

\section{APPLICATION TO GLOBAL MONTE CARLO ALGORITHMS}

The application of choice of the stochastic determinant is in Monte Carlo
algorithms. I have tested here in case of lattice Schwinger model.
To simulate the model I made new gauge field proposals using the gauge
action which consists of 10000 sweeps of local Metropolis steps followed
by a global accept/reject step to include two degenerate flavors of
fermions. Figure 1 shows the square Wilson loops as a function of the
linear size of the loop using three algorithms: 1) computing exactly
the determinant ratios, 2) using the stochastic method with $n=30$ noisy
estimators and 3) the same as in 2) but with $n=40$ noisy estimators.
This example shows that already $n=40$ estimators are enough to reach the
exact result.

\begin{figure}
\epsfxsize=3.8cm
\vspace{2cm}
\hspace{2.5cm} \epsffile[240 400 480 450]{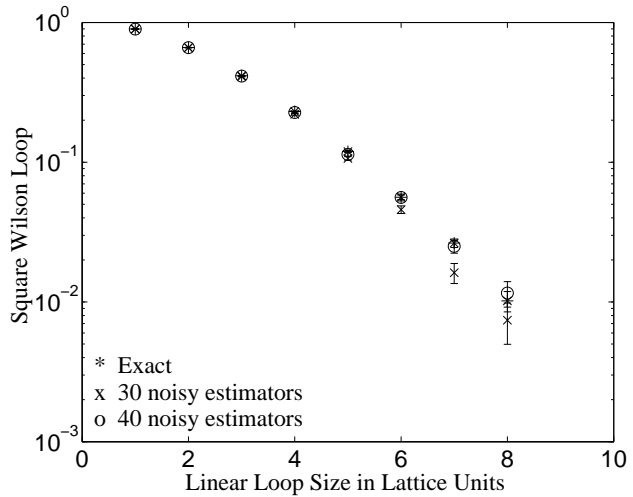}
\vspace{2.7cm}
\caption{Two flavors of Swchinger model on 
a $16x16$ lattice at $\beta = 5$
and bare quark Kogut-Susskind fermion mass $m=0.01$} 
\vspace{-.7cm}
\end{figure}

\end{document}